\documentclass{ifacconf}

\makeatletter
\let\old@ssect\@ssect 
\makeatother

\usepackage{natbib}       
\usepackage[pdftex]{graphicx}
\usepackage{algorithm}
\usepackage{algpseudocode}
\usepackage{amssymb,amsmath}
\usepackage{subfig}
\usepackage{indentfirst}
\usepackage{color}
\usepackage[hidelinks]{hyperref}
\usepackage{enumerate}
\usepackage{tikz}
\usetikzlibrary{shapes,arrows}
\usepackage{epstopdf}
\usepackage{epsfig}
\usepackage{wrapfig}
\usepackage[font=small]{caption}
\usepackage{balance}

\makeatletter
\def\@ssect#1#2#3#4#5#6{%
  \NR@gettitle{#6}
  \old@ssect{#1}{#2}{#3}{#4}{#5}{#6}
}
\makeatother

\makeatletter
\let\OldStatex\Statex
\renewcommand{\Statex}[1][3]{%
  \setlength\@tempdima{\algorithmicindent}%
  \OldStatex\hskip\dimexpr#1\@tempdima\relax}
\makeatother

\def\mf{\mathbf}
\def\mb{\mathbb}

\def\beq{\begin{equation*}}
\def\eeq{\end{equation*}}
\def\bql{\begin{equation}}
\def\eql{\end{equation}}
\def\bqn{\begin{eqnarray*}}
\def\eqn{\end{eqnarray*}}
\def\bnl{\begin{eqnarray}}
\def\enl{\end{eqnarray}}
\def\bna{\bql\begin{array}{rcl}}
\def\ena{\end{array}\eql}
\def\bnn{\beq\begin{array}{rcl}}
\def\enn{\end{array}\eeq}
\def\bma{\begin{bmatrix}}
\def\ema{\end{bmatrix}}
\def\bmx{\begin{matrix}}
\def\emx{\end{matrix}}
\def\ben{\begin{enumerate}}
\def\een{\end{enumerate}}
\def\bit{\begin{itemize}}
\def\eit{\end{itemize}}
\def\bei{\begin{itemize}}
\def\eei{\end{itemize}}
\def\bet{\begin{tabular}}
\def\eet{\end{tabular}}

\newcommand{\allcaps}[1]{\uppercase\expandafter{#1}}

\providecommand{\abs}[1]{\lvert#1\rvert} 
\providecommand{\norm}[1]{\left\|#1\right\|}

\newtheorem{remark}{Remark}
\newtheorem{definition}{Definition}
\begin{document}
\begin{frontmatter}

\title{Delay Embedded Echo-State Network: A Predictor for Partially Observed Systems}



\author[First]{Debdipta Goswami} 

\address[First]{Department of Mechanical and Aerospace Engineering, The Ohio State University, 
   Columbus, OH 43210 USA (e-mail: goswami.78@osu.edu).}

\begin{abstract}                
This paper considers the problem of data-driven prediction of partially observed systems using a recurrent neural network. While neural network based dynamic predictors perform well with full-state training data, prediction with partial observation during training phase poses a significant challenge. Here a predictor for partial observations is developed using an echo-state network (ESN) and time delay embedding of the partially observed state. The proposed method is theoretically justified with Taken's embedding theorem and strong observability of a nonlinear system. The efficacy of the proposed method is demonstrated on three systems: two synthetic datasets from chaotic dynamical systems and a set of real-time traffic data.
\end{abstract}

\begin{keyword}
Nonlinear system identification, Neural networks, Observability, Chaotic attractor, Machine learning, Reservoir computer
\end{keyword}

\end{frontmatter}

\section{Introduction}
The ongoing quest of modeling complex systems from data has motivated a wide array of machine learning techniques proving their utility in a variety of problems, e.g., classification, speech recognition \citep{Hinton2012}, board games \citep{Silver2016}, and even discovering mathematical algorithms \citep{Fawzi2022}. There has been a renewed interest in data-driven prediction of dynamical systems in the past decade where recurrent neural networks (RNN) played a central role. For example, an echo-state network (ESN) proposed by \cite{Jaeger2004} can model chaotic systems with great effect (\citealt{Lu2017}; \citealt{Pathak2018}). However, these neural network models assume that all the latent variables of the underlying system is adequately represented in training dataset. Hence, they need full-state data for training and rely on the underlying dynamic structure to approximate the state-transition map to predict a dynamical system. But in many practical cases, only a partial observation is available during the training phase. Such applications include fluid flow structure, atmospheric dynamics, and traffic data.

Neural network predictors, instead of using a handcrafted dynamic model from the physics of the system, utilize a set of training data to update a parametric surrogate model, and then employ it to predict future states. They implicitly assume that all the relevant variables are adequately represented in the training dataset and hence, require full state measurement for proper training. Although \cite{Lu2017} and \cite{Goswami2021} developed ESN-based methods to utilize sparse partial measurements for predicting unmeasured variables in the testing phase, these require full-state data for training. 

An ESN uses a reservoir of nonlinear, randomly connected neurons to process time-varying input signals. Such a reservoir with a convergence property, known to the ESN literature as echo-state property (ESP) can uniformly approximate any nonlinear fading memory filter as proved in \cite{Ortega2018}. The attractiveness of an ESN as a neural engine of a predictor is that any convergent reservoir dynamics can be tuned via output connections (also called the readout map) with minimal computing resources. Also, \cite{Tanaka2019} and \cite{Nakajima}  describe hardware implementations of the reservoir using field programmable gate arrays (FPGAs) or a photonic reservoir, thereby increasing efficiency and reducing computational overhead. It is also extended to quantum computing realm via quantum reservoir computers (QRCs) as shown in \cite{Fuji2017} and \cite{Chen2019}. \cite{Lu2017}, \cite{Goswami2021}, and \cite{Goswami2022} describe the effectiveness of ESN-based approaches for sparse estimation of chaotic systems and traffic network prediction.

This paper develops an ESN-based predictor for systems with partial measurements available for network training. The proposed method utilizes the universal approximation property of a fading memory ESN \citep{Ortega2018} coupled with Taken's embedding theorem \citep{Takens1981} to determine a dynamic map that predict the future values of the partial observation. \cite{Takens1981} showed that a time series of typical scalar measurements can faithfully reconstruct the attractor set of a chaotic dynamical system. This result motivates the utilization of a delay embedded partial measurements to train the ESN as a dynamic map for next step prediction. It is further proved for a general nonlinear system that a finite time observability condition on an open set is sufficient for existence of such a dynamic map.

The contribution of this paper are (1) providing a data-driven predictor for partial observations from a higher dimensional nonlinear systems via ESN; (2) utilization of Taken's embedding theorem and strong observability condition to guarantee the existence of such dynamic predictor; (3) application of the prediction method on a real set of mobility data in order to forecast traffic volume in a road network.

This paper is organized as follows. Section 2 provides a brief overview of the echo-state network (ESN). Section 3 presents the ESN algorithm with delay embedded input and provide theoretical justification of the algorithm. Section 4 illustrates the applications to three different problems: two synthetic data streams generated by chaotic nonlinear systems and one real set of traffic sensor data. An ablation study with different embedding dimension is also provided. Section 5 concludes the manuscript and discusses ongoing and future work.

%
%

\section{Echo-State Networks: Dynamical System Predictor}
\begin{figure*}[t]
\centering 
\subfloat[]{\includegraphics[trim=0cm 0cm 0cm 0cm, clip=true, width=0.33\textwidth]{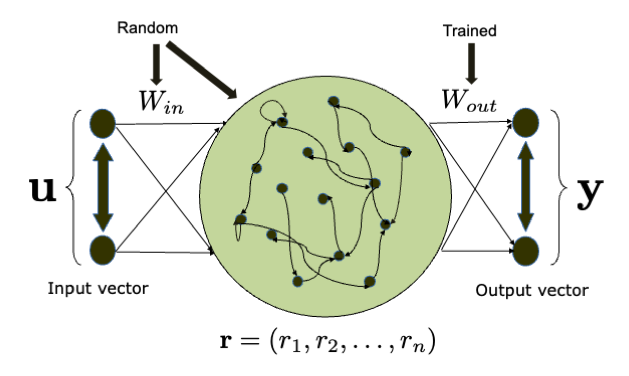}}
\subfloat[]{\includegraphics[trim=0cm 0cm 0cm 0cm, clip=true, width=0.33\textwidth]{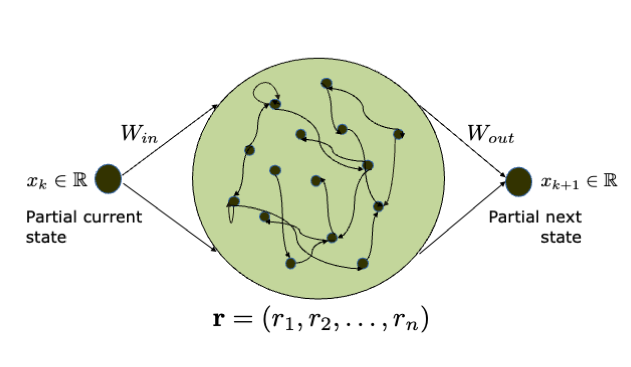}}
\subfloat[]{\includegraphics[trim=0cm 0cm 0cm 0cm, clip=true, width=0.33\textwidth]{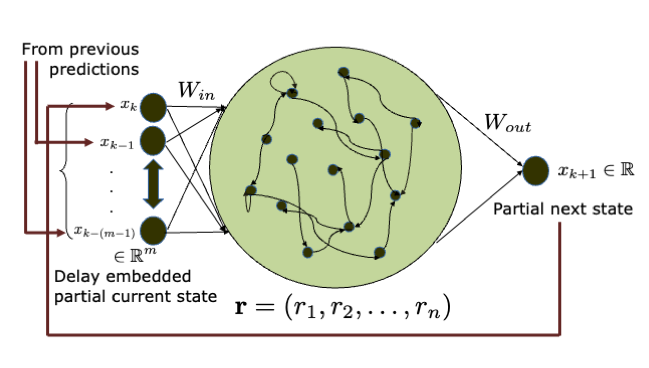}}
\caption{Architecture of an EnKF-ESN: (a) the basic ESN, (b) an ESN for scalar partial state prediction, and (c) a delay-embedded ESN for partial state prediction } \label{Fig: Reservoir}
\end{figure*}

Echo-state networks (ESN) are a special kind of recurrent neural network used for the prediction of dynamical systems and time-series. It consists of a large, randomly connected reservoir of neurons driven by the input signal (Fig.~\ref{Fig: Reservoir}.(a)). The nonlinear response signals thus induced in the neurons are then linearly combined to match a desired output signal. This technique is also known as reservoir computer (RC) \citep{Maass2004}. An ESN consists of an input layer $\mf{u}\in\mb{R}^m$, coupled through input coupling matrix $W_{in}\in \mb{R}^{n\times m}$ with a recurrent nonlinear reservoir $\mf{r} \in \mathbb{R}^n$. The output $\mathbf{y}\in \mathbb{R}^p$ is generated from $n$ neurons of the reservoir via a readout matrix $W_{out}\in \mb{R}^{n\times p}$. The reservoir network evolves nonlinearly in following  fashion (\citealt{Maass2004}; \citealt{Goswami2021})
\bql
\mf{r}(t+\Delta t)=(1-\alpha)\mf{r}(t) + \alpha\psi(W\mf{r}(t)+W_{in}\mf{u}(t)).
\eql
The time-step $\Delta t$ is chosen according to the sampling interval of the training data. The leakage rate parameter $\alpha \in (0,1]$ slows down the evolution of the reservoir as $\alpha \rightarrow 0$. The nonlinear activation function $\psi$ is usually a static nonlinear function, e.g., $\tanh(\cdot)$ or a logistic function. The output depends linearly on the reservoir states (\cite{Maass2004}, \cite{Goswami2021}), i.e.,
\bql
\mf{y}(t)=W_{out}\mf{r}(t).
\eql
The weights $W_{in}$ and $W$ are initially randomly drawn from according to a random graph model and then held fixed. $W_{in}$ can contain a column to inherently supply a bias with unit input. The weight $W_{out}$ is adjusted during the training process. The reservoir weight matrix $W$ is usually kept sparse for computational efficiency.  

An ESN is trained by driving it with an input sequence $\{\mf{u}(1),\ldots,\mf{u}(N)\}$ that yields a sequence of reservoir states $\{\mf{r}(1),\ldots,\mf{r}(N)\}$. The reservoir states are stored in a matrix $\mf{R}=[\mf{r}(t_1),\ldots,\mf{r}(t_N)]$. The correct outputs $\{\mf{y}(1),\ldots,\mf{y}(N)\}$, which are part of the training data, are also arranged in a matrix $\mf{Y}=[\mf{y}(1),\ldots,\mf{y}(N)]$.  The training is carried out by a linear regression with Tikhonov regularization as follows \citep{Jaeger2004}:
\bql
W_{out} = \mf{Y}\mf{R}^T(\mf{R}\mf{R}^T + \beta\mf{I})^{-1},
\eql
where $\beta>0$ is a regularization parameter that ensures non-singularity. 
\begin{remark}
An ESN is a universal approximator, i.e., it can realize every nonlinear operator with bounded memory arbitrarily accurately, if it satisfies the echo state property (ESP) as explained in \cite{Jaeger2004}. The ESP states that the reservoir will asymptotically wash out any information from the initial conditions. For the $\tanh(\cdot)$ activation function, \cite{Jaeger2004} empirically observed that the ESP holds for any input if the spectral radius of $W$ is smaller than unity. To ensure this condition, $W$ is normalized by its spectral radius. 
\end{remark}

\section{Delay Embedded Echo-State Network: A Predictor for Partial Scalar Observation}
An ESN can be trained to predict a time-series $\{\mf{x}_{i}\in \mb{R}^d:i\in\mb{N}\}$ generated by a dynamical system by setting $\mf{u}(t)$ and $\mf{y}(t)$ as the current and next state value (i.e., $\mf{x}_{k}$ and $\mf{x}_{k+1}$) respectively. The network is trained for a certain training length $N$ of the time-series data $\{\mf{x}_{i},i=1,\ldots,N\}$, and then can run freely by feeding the output $\mf{y}_{k}$ back to the input $\mf{u}_{{k+1}}$ (i.e., $\mf{u}_{{k+1}}\gets \mf{y}_{k}$) of the reservoir. In this case, both $\mf{u}$ and $\mf{y}$ have the same dimension $d$ as that of the time-series data. 

An ESN proves to be a powerful tool for dynamical systems prediction when trained with full-state data as demonstrated in \cite{Lu2017} and \cite{Pathak2018}. \cite{Goswami2021} improves its performance by assimilating partial observations during the testing phase through an ensemble Kalman filter. However, a significant challenge is posed when only a partial scalar measurement $\{{x}_{i}\in \mb{R}:i\in\mb{N}\}$ of the state is available during the training phase. A simple approach is to treat the scalar measurements as a separate time-series and use it train an ESN Fig. \ref{Fig: Reservoir} (b). The trained ESN can then be used to predict the future values of the scalar time series $\{{x}_{i}\in \mb{R}:i=N+1,\ldots\}$. But this simplistic approach disregards the fact that the scalar time series comes from a high dimensional system and the next step prediction $x_{k+1}$ of the scalar observable might depend on not only $x_k$ but its previous values $x_{k-(m-1)},\ldots,x_k$ via higher dimensional embedding. 

This paper proposes an alternative method of partial state prediction by utilizing time delay embedding in conjunction with an ESN. This approach is inspired by \cite{Takens1981} and \cite{Sauer1991} which show that the time-delay embedding of a scalar measurement time-series can diffeomorphically reconstruct the strange attractor of the original dynamics. This result is famously stated as Taken's embedding theorem:
\begin{thm}\label{Thm: Takens}
Let a discrete-time dynamical system $\mf{x}_{k+1} = f(\mf{x}_k)$ is given by a smooth map $f:\mb{X}\rightarrow \mb{X}$ on a $d$ dimensional manifold $\mb{X}\in\mb{R}^d$. Assume that the dynamics evolve on a strange attractor $A$ with a box-counting dynamics $d_A$. Then $A$ can be embedded in $\mb{R}^m$ with $m>2d_A$, i.e., $\exists$ a diffeomorphism $\phi : A \rightarrow \mb{R}^m$ such that the derivative of $\phi$ is full rank. Moreover, let $h:\mb{X}\rightarrow \mb{R}$ be a smooth scalar observation function with full rank derivative and no special symmetry in the component. Then the function $e:\mb{X}\rightarrow \mb{R}^m$ defined as 
\[e(\mf{x}) \triangleq \left( h(\mf{x}), h(f(\mf{x})),\ldots,h(f^{m-1}\mf{x})\right)\]
is an embedding of $A$ in $\mb{R}^k$.
\end{thm}
\begin{pf}
See \cite{Takens1981}. \hfill $\blacksquare$
\end{pf}

\begin{figure}[h]
\centering 
\includegraphics[trim=0cm 0cm 0cm 0cm, clip=true, width=0.5\textwidth]{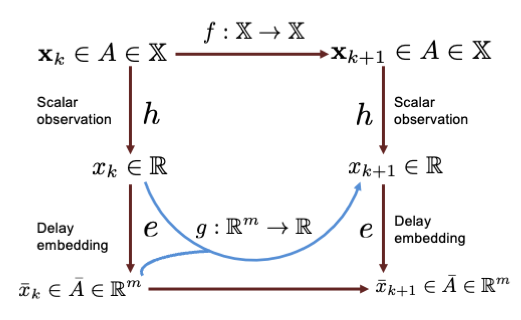}
\caption{Schematic of the Takens embedding theorem \citep{Vlachos2008}}\label{Fig: TakensSchematic}
\end{figure}
The embedding is demonstrated in Fig. \ref{Fig: TakensSchematic}. Since the scalar measurements $\{x_i:i\in\mb{N}\}$ can be delay embedded to an attractor $\bar{A}$ diffeomorphic to $A$, there is a map $g:\mb{R}^m\rightarrow\mb{R}$ to the next step scalar observation $x_{k+1}$ from the time-delay embedding of its previous values $\{x_k,\ldots,x_{k-m+1}\}$. This result, coupled with the universal approximation property of an ESN \cite{Ortega2018}, provides a strong basis to use delay embedding for the input layer in order to predict the partial scalar observation (Fig. \ref{Fig: Reservoir}(c)). Algorithm \ref{Alg: DelayESN} presents the procedure for training an ESN using delay-embedded scalar measurements. The inputs are taken as vectors $\bar{x}_k = [x_k,\ldots,x_{k-m+1}]^T$ of $m$ delayed observations. The ESN is trained with next step scalar observation $x_{k+1}$ as outputs.

\begin{remark}
The condition of full rank derivative and no special symmetry on the observation function $h$ is important and closely relates to the notion of observability of a nonlinear system. 
\end{remark}

\begin{remark}
The embedding dimension $m$, as specified in Theorem \ref{Thm: Takens}, is at most $2d + 1$, but it can be often less in reality. For example, Lorenz system yields a theoretical value of $m=5$ because its strange attractor has a box-counting dimension $d_A=2.06\pm0.01$.  For a delay-embedded ESN (Fig. \ref{Fig: Reservoir}(c)), $m$ is a hyperparameter that needs to be tuned for the best performance. 
\end{remark}

\begin{remark}
While Taken's embedding theorem is valid for systems with strange attractors, the delay embedded ESN can be used for any strongly observable nonlinear system on its domain of observability as will be shown next. This method is also suited for quasi-periodic systems as demonstrated in this paper.
\end{remark}

\begin{algorithm}
\caption{Training a delay-embedded ESN}\label{Alg: DelayESN}
\textbf{Input:} Training scalar measurements $\{x(1),\ldots,x(N)\}$\\
\textbf{Hyperparameters:} Delay embedding dimension $m$, Training length $N$, leaking rate $\alpha$, regularization parameter $\beta$, reservoir connection probability $p\in(0,1)$, reservoir size $n$, activation $\psi$\\
\textbf{Output:} $W_{in}$, $W$, $W_{out}$
\begin{algorithmic}[1]
\Procedure {Train}{ $\{x(1),\ldots,x(N)\}$; $m$, $\alpha$, $\beta$, $p$, $n$, $\psi$}
\State Generate $W\sim G(n,p)$ \Comment{Adjacency matrix of an \Statex Erd\"os-Renyi random graph}
\State Generate $W_{in} \sim \operatorname{uniform}(-0.5, 0.5)^{n\times m}$ \Statex random matrix
\State $\bar{x}(i) \gets [x(i),x(i+1),\ldots,x(i+m-1)]^T$ \Statex $\forall$ $i\in{1,\ldots,N-m}$ \Comment{Delay embedding}
\State $\mf{Y} \gets [x(m+1), x(m+2), \ldots, x(N)]$ \Statex\Comment{Arrange outputs}
\State $\mf{r}_1 \gets \mf{0}_n$ \Comment{Initialize reservoir}
\For{$i=1$ to $N-m$}
\State $\mf{r}(i+1) \gets (1-\alpha)\mf{r}(i) + \alpha\psi(W\mf{r}(i)+W_{in}\bar{x}(i))$
\EndFor
\State $\mf{R} \gets [\mf{r}(1),\ldots,\mf{r}(N-m)]$ \Statex\Comment{Arrange reservoir states}
\State $W_{out} \gets \mf{Y}\mf{R}^T(\mf{R}\mf{R}^T + \beta\mf{I})^{-1}$ \Statex\Comment{Train output weights}
\EndProcedure
\end{algorithmic}
\end{algorithm}

A time-series of partial observation $\{{x}_1,\ldots,{x}_N\} = \{h(\mf{x}_0), h^2(f(\mf{x}_0)),\ldots,h(f^{N-1}(\mf{x}_0)\}$ from a dynamical system $\mf{x}_{k+1} = f(\mf{x}_k)$ with an observation function $h(\cdot)$ is able to predict the next step by training an universal functional approximator if the observability condition is satisfied as stated below.

\begin{definition} \citep{Nijmeijer1982} \label{Def: Observability}
Consider an autonomous discrete-time nonlinear system
\bnl \label{Eq: System}
\mf{x}_{k+1} &=& f(\mf{x}_k) \\\nonumber
y_k &=& h(\mf{x}_k), \, k\in\mb{N}\cup \{0\},
\enl
where $f:\mb{X}\subset\mb{R}^d \rightarrow \mb{X}$ and $h:\mb{X}\rightarrow\mb{R}$ are smooth functions defined on an open subset $\mb{X} \subset \mb{R}^d$. The system \eqref{Eq: System} is said to be \emph{strongly observable}, or \emph{finite-time observable} if for any $\mf{x}$, $\tilde{\mf{x}} \in \mb{X}$, \[\begin{bmatrix}
h(\mf{x})\\ h(f(\mf{x}))\\\vdots\\h(f^{d-1}(\mf{x})) \end{bmatrix} = \begin{bmatrix}h(\tilde{\mf{x}})\\ h(f(\tilde{\mf{x}}))\\\vdots\\h(f^{d-1}(\tilde{\mf{x}})\end{bmatrix}\] implies $\mf{x} = \tilde{\mf{x}}$. 
\end{definition}

\begin{table*}[t]
\captionsetup{width=.75\textwidth}
\caption{ESN hyperparameters}
\begin{center}
 \label{tb:hyparam}
\begin{tabular}{lccc}
\hline
Hyperparameter & &Value &  \\
 & Lorenz system \eqref{Eq: Lorenz} & R\"{o}ssler system \eqref{Eq: Rossler} & Traffic Volume \\\hline
Reservoir size $n$ & $500$ & $500$ & $4000$\\
Reservoir connection probability $p$ & $0.01$ & $0.01$ & $0.01$ \\
Training length $N$ & $1000$ & $1000$ & $1000$ \\ 
Activation $\psi(\cdot)$ & $\tanh(\cdot)$ & $\tanh(\cdot)$ & $\tanh(\cdot)$\\
Leaking rate $\alpha$ & $0.3$ & $0.3$ & $0.7$\\
Regularization $\beta$ & $10^{-6}$ & $10^{-6}$ & $10^{-6}$\\\hline
\end{tabular}
\end{center}
\end{table*}

For this paper, we assume the observation $y_k$ are partial state $x_k$.
\begin{thm}\label{Thm: Observability}
Consider a scalar time series  $\{x_i:i\in\mb{N}\cup \{0\}\}$ observed from a nonlinear dynamics \eqref{Eq: System} with a full state history $\{\mf{x}_i: i\in\mb{N}\cup \{0\}\}$, and observation function $h(\cdot)$ such that $y_k = x_k = h(\mf{x}_k)$. If the system \eqref{Eq: System} is strongly observable, then there is a continuous map $g:\mb{R}^d \rightarrow \mb{R}$ from a delay-embedded partial state observation $\bar{x}_k = [x_k,\ldots,x_{k-d+1}]$ to the next step $x_{k+1}$. Moreover, there exists an ESN, properly chosen, that can approximate $g$ arbitrarily accurately.
\end{thm}

\begin{pf}
Let $\mb{Y}\subset \mb{R}^d$ be the set of all possible $d$-step observation sequence, i.e., $\mb{Y} = [h, h\circ f, \ldots, h\circ f^{d-1}](\mb{X})$. The strong observability condition in Definition \ref{Def: Observability} states the existence of a bijection $H(\cdot)\triangleq [h, h\circ f, \ldots, h\circ f^{d-1}](\cdot)$ from $\mb{X}$ to $\mb{Y}$. Now since $H$ is smooth (i.e., at least continuously differentiable) by construction on an open set $\mb{X}$, $\mb{Y} = H(\mb{X})$ is also open in $\mb{R}^d$ and $H$ is a homeomorphism between $\mb{X}$ and $\mb{Y}$. Hence, $H$ has a continuous inverse $G:\mb{Y}\rightarrow \mb{X}$ which maps the delay-embedded observation sequence $\bar{x}_k = [x_k,\ldots,x_{k-d+1}]$ to $\mf{x}_{k-d+1}$. Now according to the dynamics \eqref{Eq: System}, $x_{k+1} = h\circ f^d(\mf{x}_{k-d+1})$ with smooth $h\circ f^d$. Therefore, we can construct a continuous function $g:\mb{Y}\subset \mb{R}^d \rightarrow \mb{R}$ such that $g \triangleq  h\circ f^d \circ G$. The last part of the theorem is a direct result of Theorem 4.1 in \cite{Ortega2018} which proves the universality of an ESN. \hfill $\blacksquare$
\end{pf}

\begin{remark}
Although \emph{strong observability} is assumed in Theorem \ref{Thm: Observability}, a weaker condition of \emph{strong local observability} \citep{Nijmeijer1982} suffices for most nonlinear systems in practice. A system \eqref{Eq: System} is \emph{strongly locally observable} at a point $\mf{x} \in \mb{X}$ if the strong observability condition is satisfied in a neighborhood $U$ of $\mf{x}$.  
\end{remark}

\begin{remark}
With the strong observability assumption, the embedding dimension $m$ should equal the system state dimension $d$. But in practical implementations, $m$ needs to be tuned to get the best performance.
\end{remark}

\section{Numerical Performance}
This section illustrates the performance and ablation study of a delay embedded ESN on three prediction problems from partial state data. The first two are time-series generated by chaotic dynamical systems and the last one is a real-time traffic flow data obtained by Numina sensor nodes \citep{Numina} installed on the University of Maryland campus.

\subsection{Lorenz System}
\begin{figure}[t]
\centering 
\subfloat[]{\includegraphics[trim=2cm 0cm 0cm 0cm, clip=true, width=0.5\textwidth]{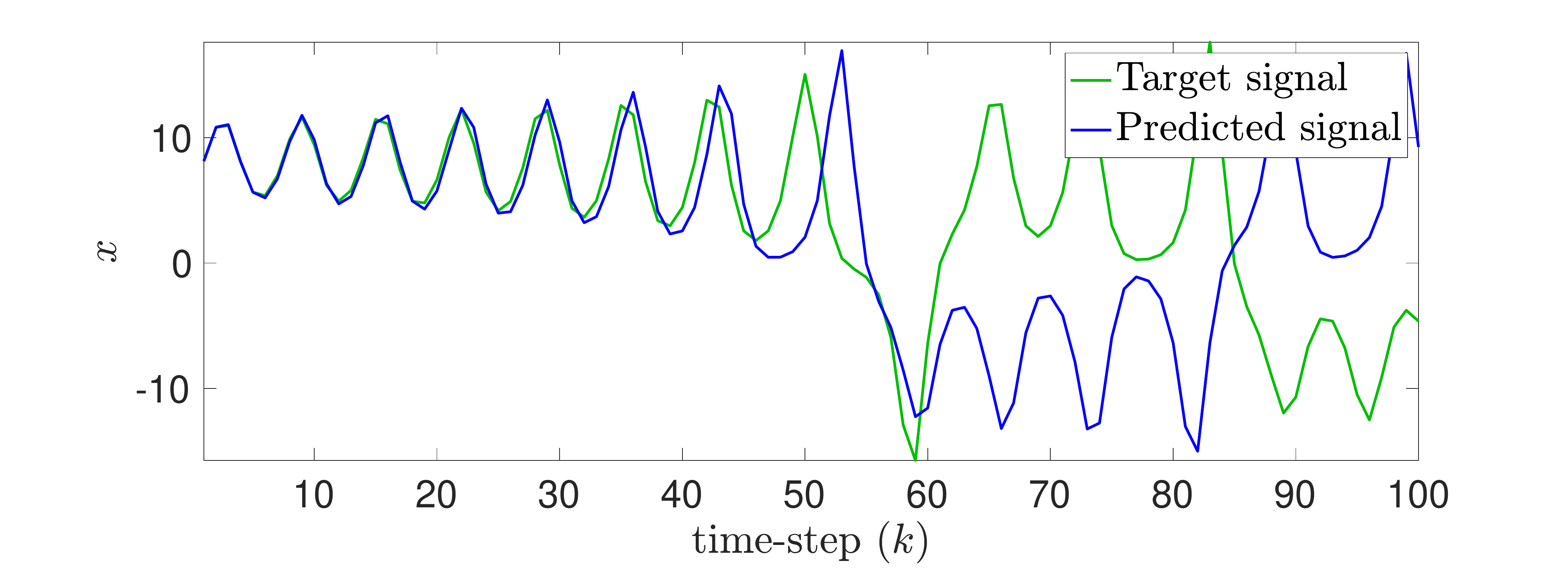}}\\
\subfloat[]{\includegraphics[trim=2cm 0cm 0cm 0cm, clip=true, width=0.5\textwidth]{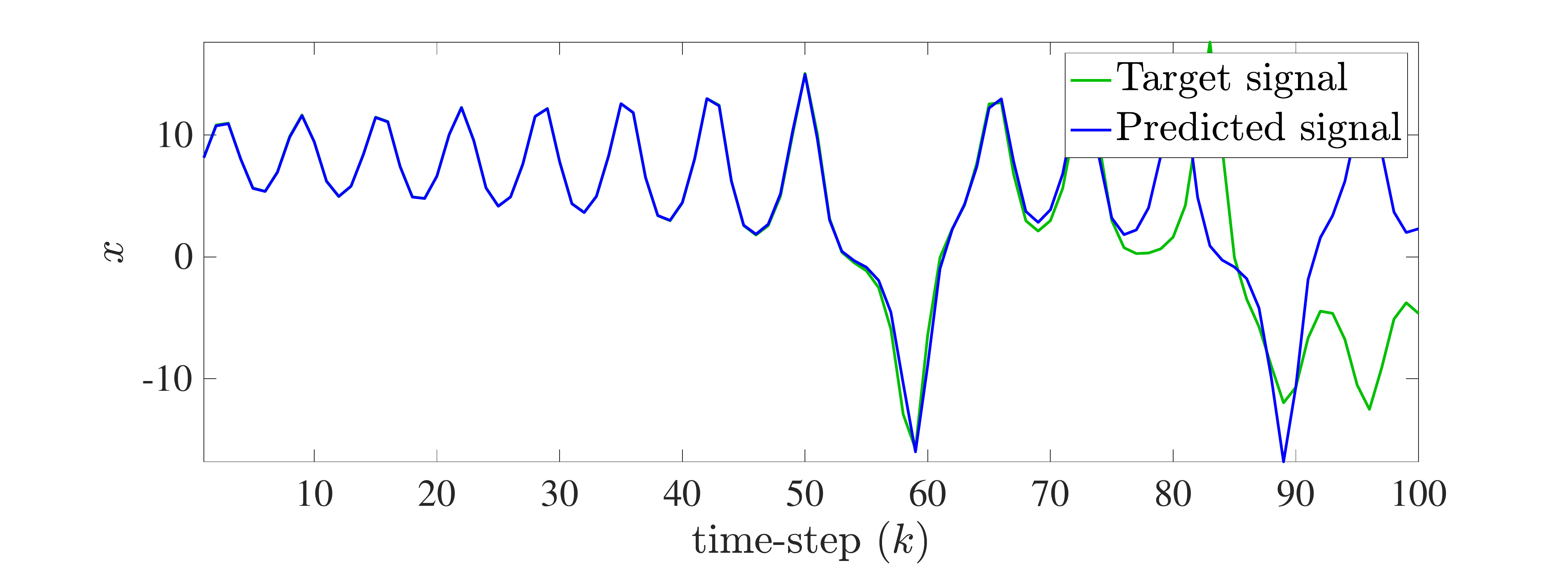}}
\caption{Estimation of the partially observed time-series $x(k)$ from Lorenz system \eqref{Eq: Lorenz} (a) true and estimated signal with no delay embedding ($m=1$), (b)  true and estimated signal with 5 dimensional delay embedding ($m=5$)} \label{Fig: Lorenz}
\end{figure}

\begin{figure}[t]
\centering 
\subfloat[]{\includegraphics[trim=0cm 0cm 0cm 0cm, clip=true, width=0.25\textwidth]{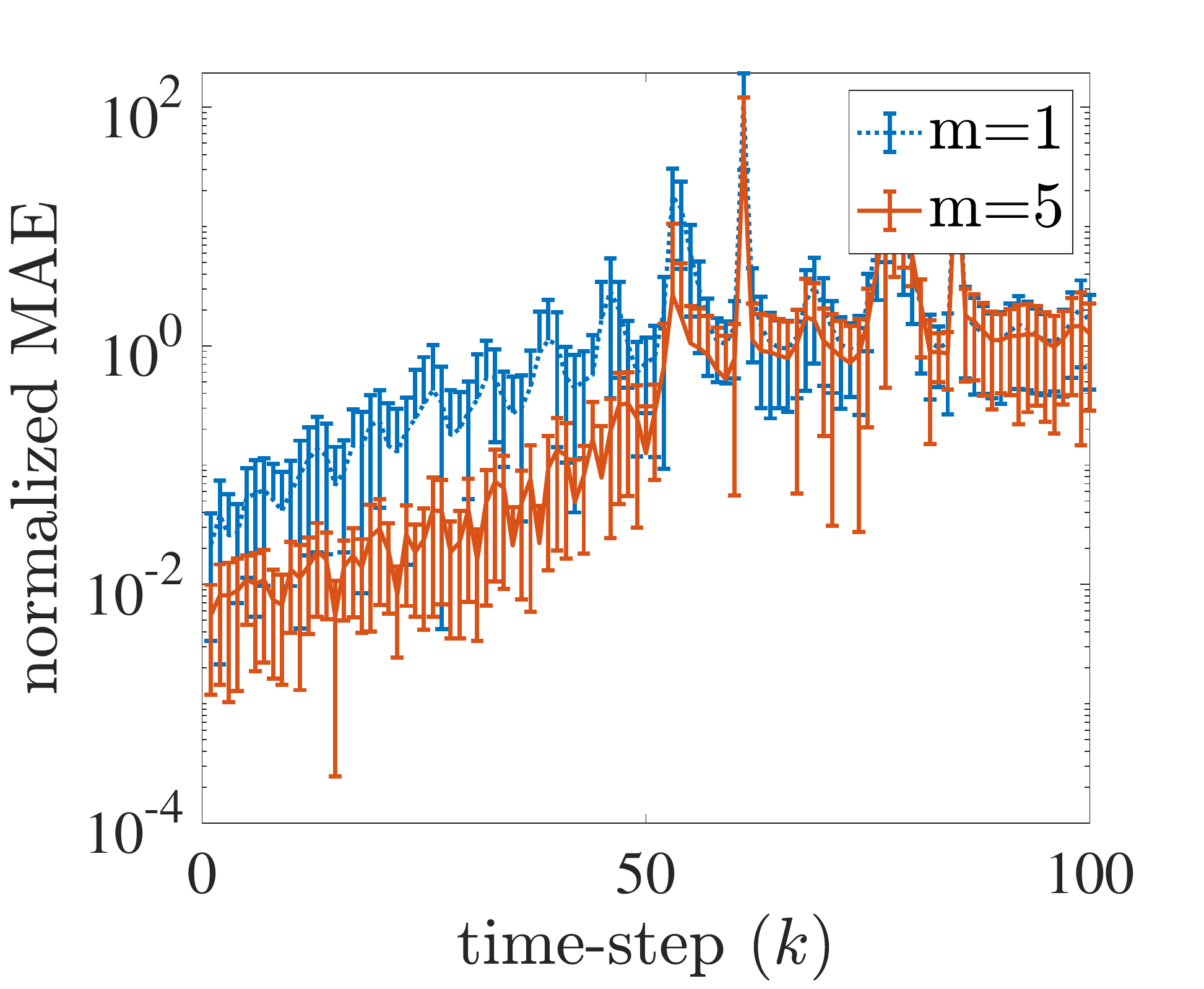}}
\subfloat[]{\includegraphics[trim=0cm 0cm 0cm 0cm, clip=true, width=0.25\textwidth]{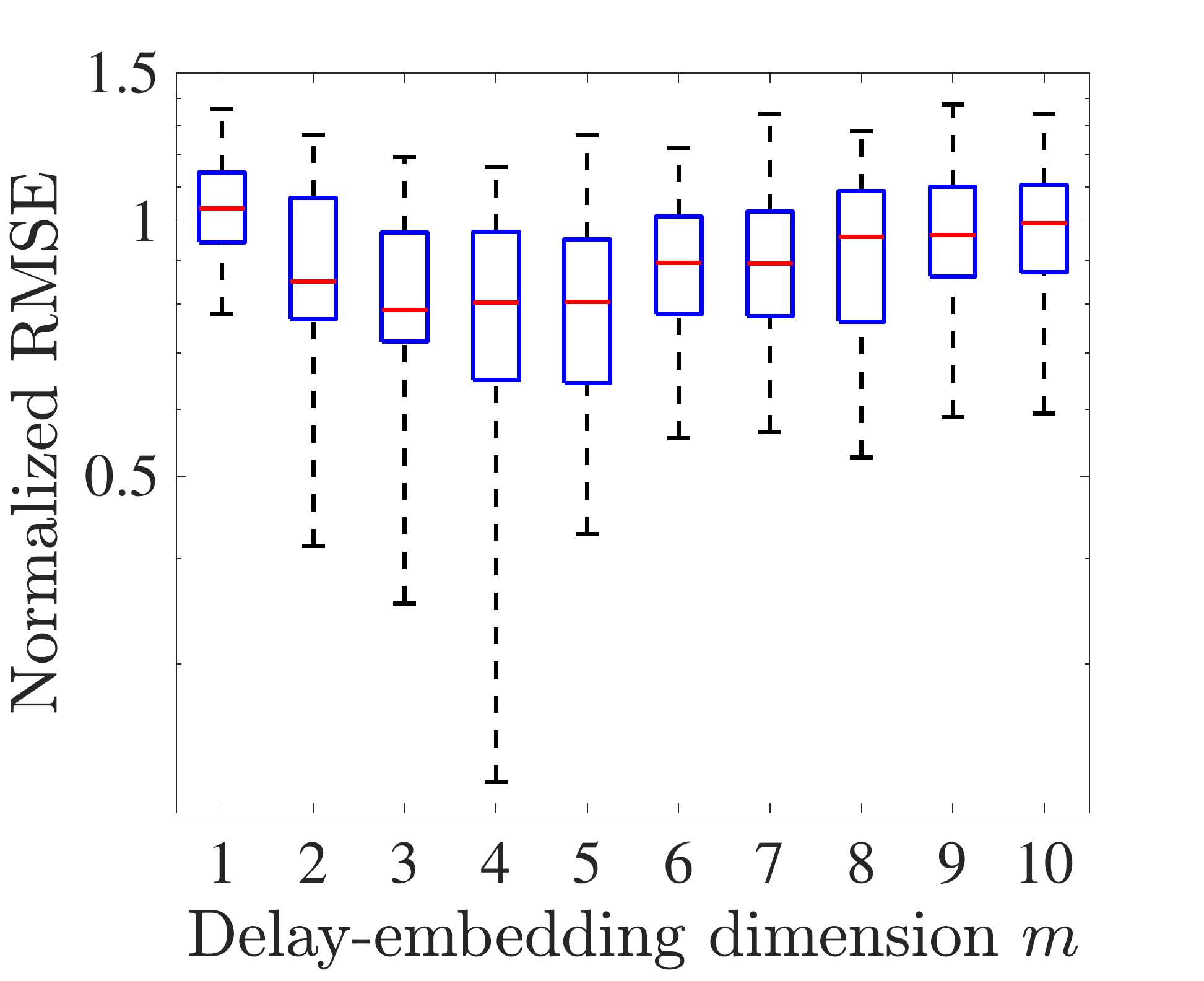}}
\caption{Error profile of partially observed Lorenz time-series estimation: (a) NMAE with time for different embedding dimension $m$, (b) NRMSE with different embedding dimension $m$} \label{Fig: LorenzError}
\end{figure}

The delay-embedded ESN is tested on a time-series generated by the Lorenz system:
\bnl\label{Eq: Lorenz}
\dot{x} &=& \sigma(y-x)\\\nonumber
\dot{y} &=& x(\rho-z) -y \\\nonumber
\dot{z} &=& xy - \beta z,
\enl
where $\sigma=10$, $\rho=28$, and $\beta=8/3$ produces chaotic behavior. Only the first state $x(k)$ with $\Delta t = 0.1$ is observed. Table \ref{tb:hyparam} lists the hyperparameters used to train the ESN. The performance of an ESN with and without delay embedding is depicted in Fig.~\ref{Fig: Lorenz}. Fig.~\ref{Fig: LorenzError} provides a detailed error profile for different embedding dimensions $m$. The normalized mean absolute error (NMAE) between true and predicted scalar value ($x(k)$ and $\hat{x}(k)$ resp.) is given by $\abs{x(k)-\hat{x}(k)}/\abs{x(k)}$ and plotted with time in Fig.~\ref{Fig: LorenzError}(a) for $m=1$ and $5$. The normalized root mean square error (NRMSE) between the true sequence $\{x(i): i=1,\ldots, l\}$ and the predicted sequence $\{\hat{x}(i): i=1,\ldots, l\}$ is given by
\bql
\operatorname{NRMSE}(x,\hat{x}) = \sqrt{\dfrac{\sum\limits_{i=1}^l \norm{x(i)-\hat{x}(i)}^2}{\sum\limits_{i=1}^l \norm{x(i)}^2}},
\eql
where $l$ is the prediction length. The prediction NRMSEs for different delay embeddings over 50 independent Monte-Carlo trials are plotted in Fig.~\ref{Fig: LorenzError}(b). As we can see, the error median is lowest with $m=5$ as predicted by Taken's theorem.

\subsection{R\"{o}ssler System}
\begin{figure}[t]
\centering 
\subfloat[]{\includegraphics[trim=2cm 0cm 0cm 0cm, clip=true, width=0.5\textwidth]{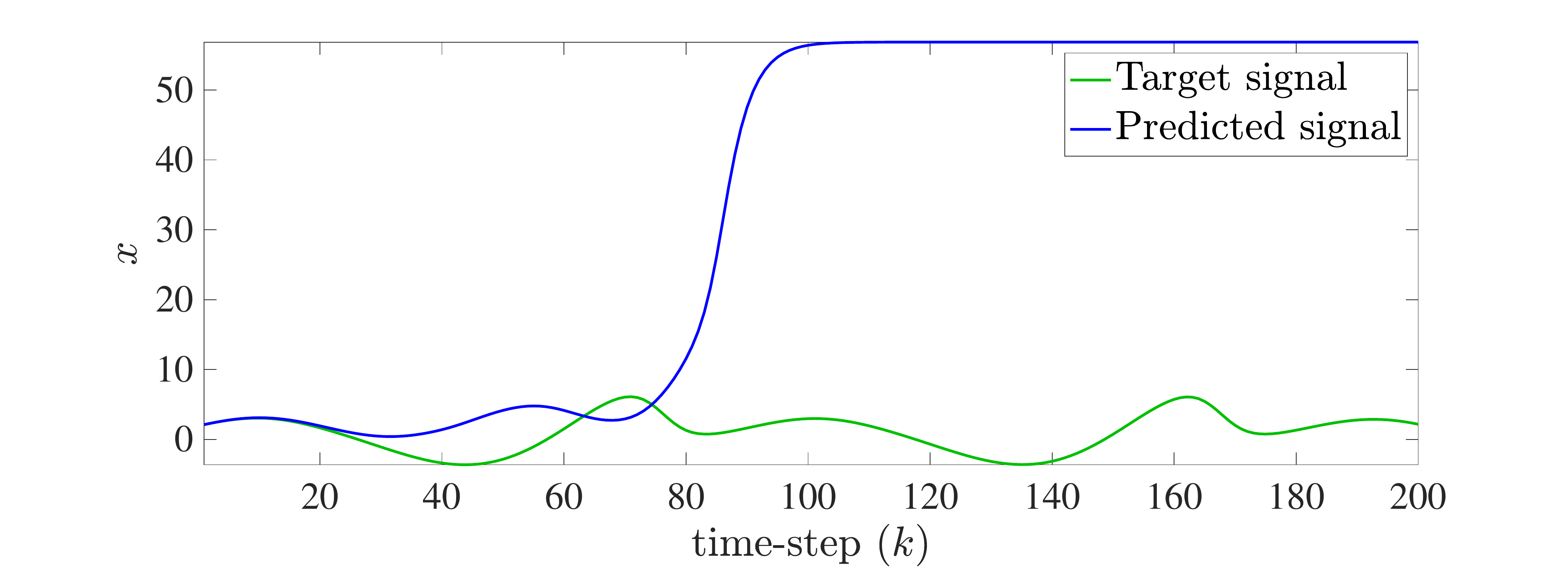}}\\
\subfloat[]{\includegraphics[trim=2cm 0cm 0cm 0cm, clip=true, width=0.5\textwidth]{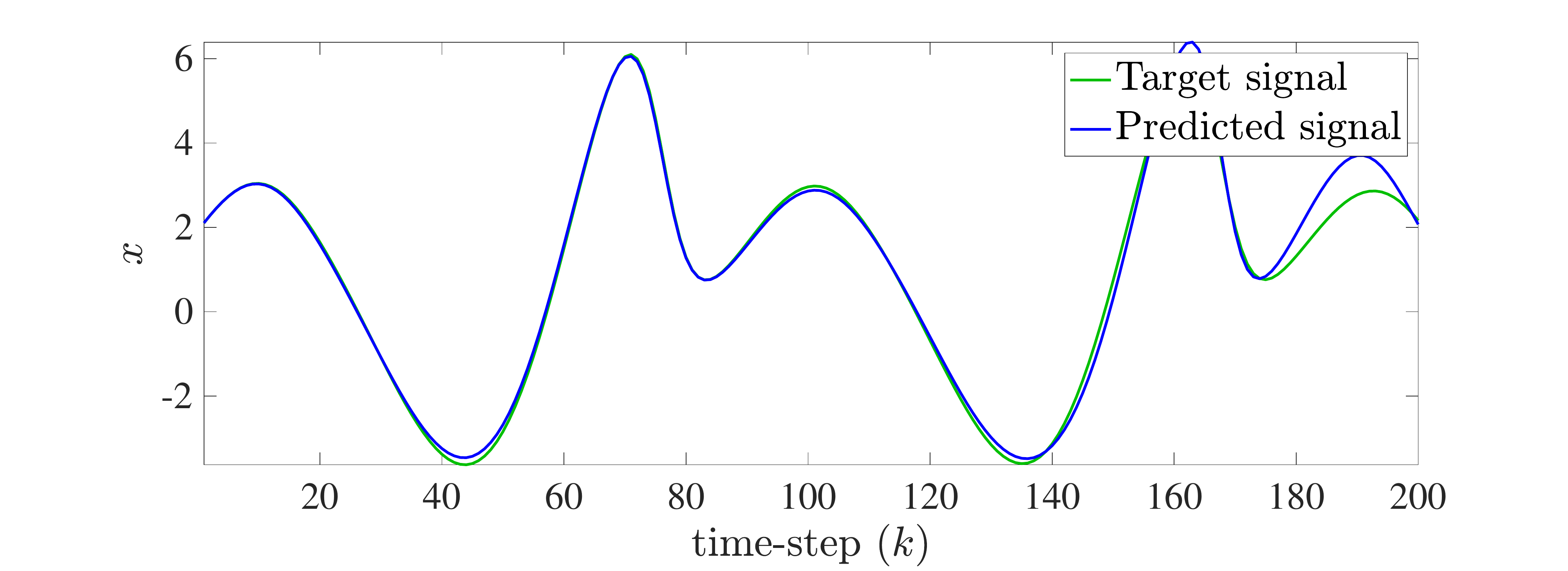}}
\caption{Estimation of the partially observed time-series $x(k)$ from R\"{o}ssler system \eqref{Eq: Rossler}: (a) true and estimated signal with no delay embedding ($m=1$), (b)  true and estimated signal with 5 dimensional delay embedding ($m=5$)} \label{Fig: Rossler}
\end{figure}

\begin{figure}[t]
\centering 
\subfloat[]{\includegraphics[trim=0cm 0cm 0cm 0cm, clip=true, width=0.25\textwidth]{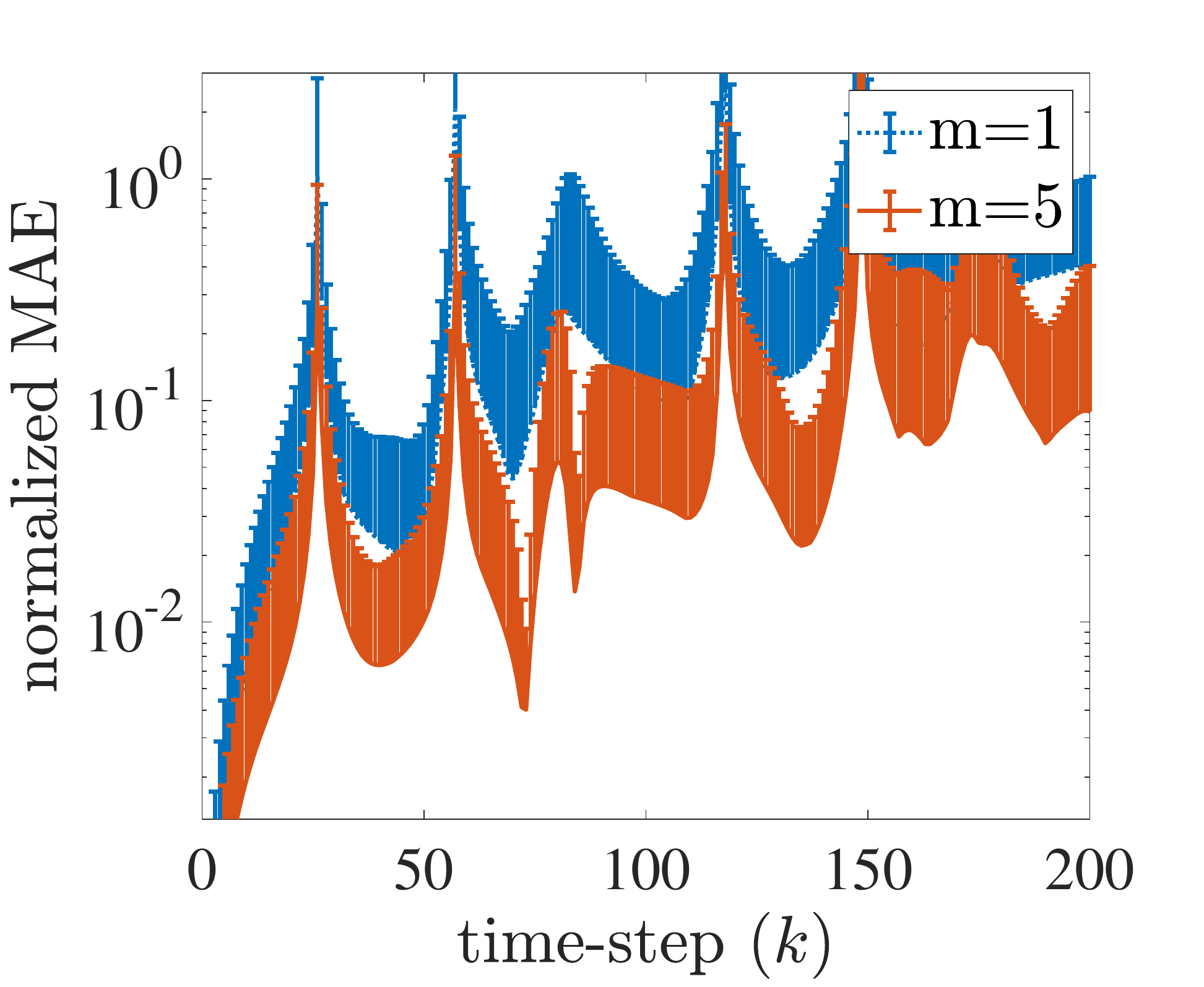}}
\subfloat[]{\includegraphics[trim=0cm 0cm 0cm 0cm, clip=true, width=0.25\textwidth]{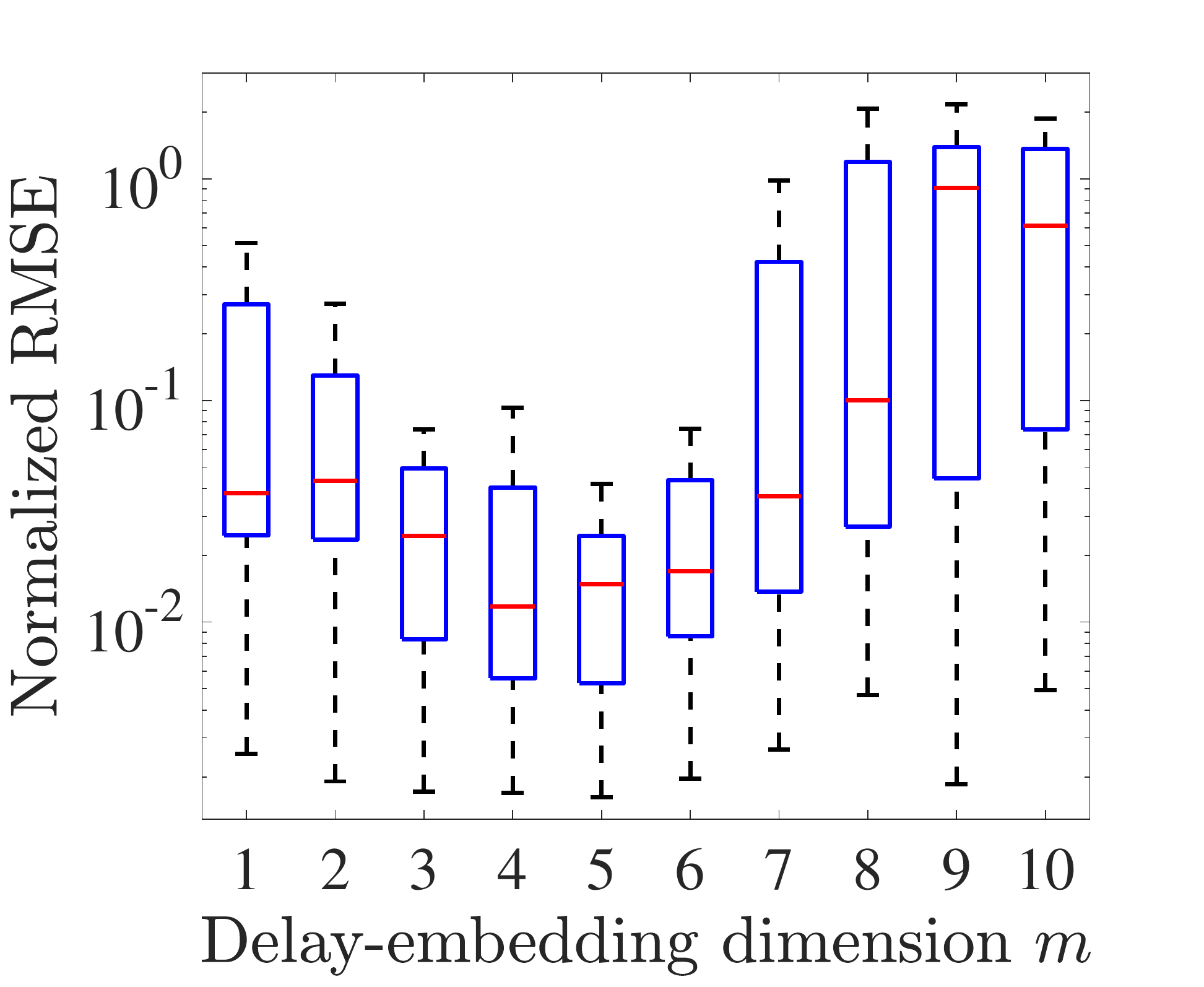}}
\caption{Error profile of partially observed R\"{o}ssler time-series estimation: (a) NMAE with time, (b) NRMSE with different embedding dimension $m$} \label{Fig: RosslerError}
\end{figure}

Next, the delay-embedded ESN is utilized to predict the partial state measurement generated by the R\"{o}ssler system described in \cite{Rossler1976}:
\bnl\label{Eq: Rossler}
\dot{x} &=& -y-x\\\nonumber
\dot{y} &=& x + ay \\\nonumber
\dot{z} &=& b + z(x-c),
\enl
with $a=0.5$, $b=2$, and $c=4$ to produce chaotic behavior. Similar to the Lorenz system example, only $x(k)$ with $\Delta t = 0.1$ is observed. Table \ref{tb:hyparam} lists the hyperparameters used to train the ESN. The performance of an ESN with and without delay embedding is depicted in Fig.~\ref{Fig: Rossler}. Fig.~\ref{Fig: RosslerError}(a) provides a the normalized MAE with time with different values of the embedding dimension $m$.  Fig.~\ref{Fig: RosslerError}(b) plots the overall NRMSE for different $m$. Here also, $m=5$ yields the best performance and almost an order of magnitude improvement in error than its counterpart with no delay embedding, i.e., $m=1$. The results are generated by 50 independent Monte-Carlo trials for training and testing the ESNs.


\subsection{Prediction of Traffic Volume on an Intersection of a Road Network}
\begin{figure}[t]
\centering 
\subfloat[]{\includegraphics[trim=1cm 0.5cm 2cm 0.2cm, clip=true, width=0.25\textwidth]{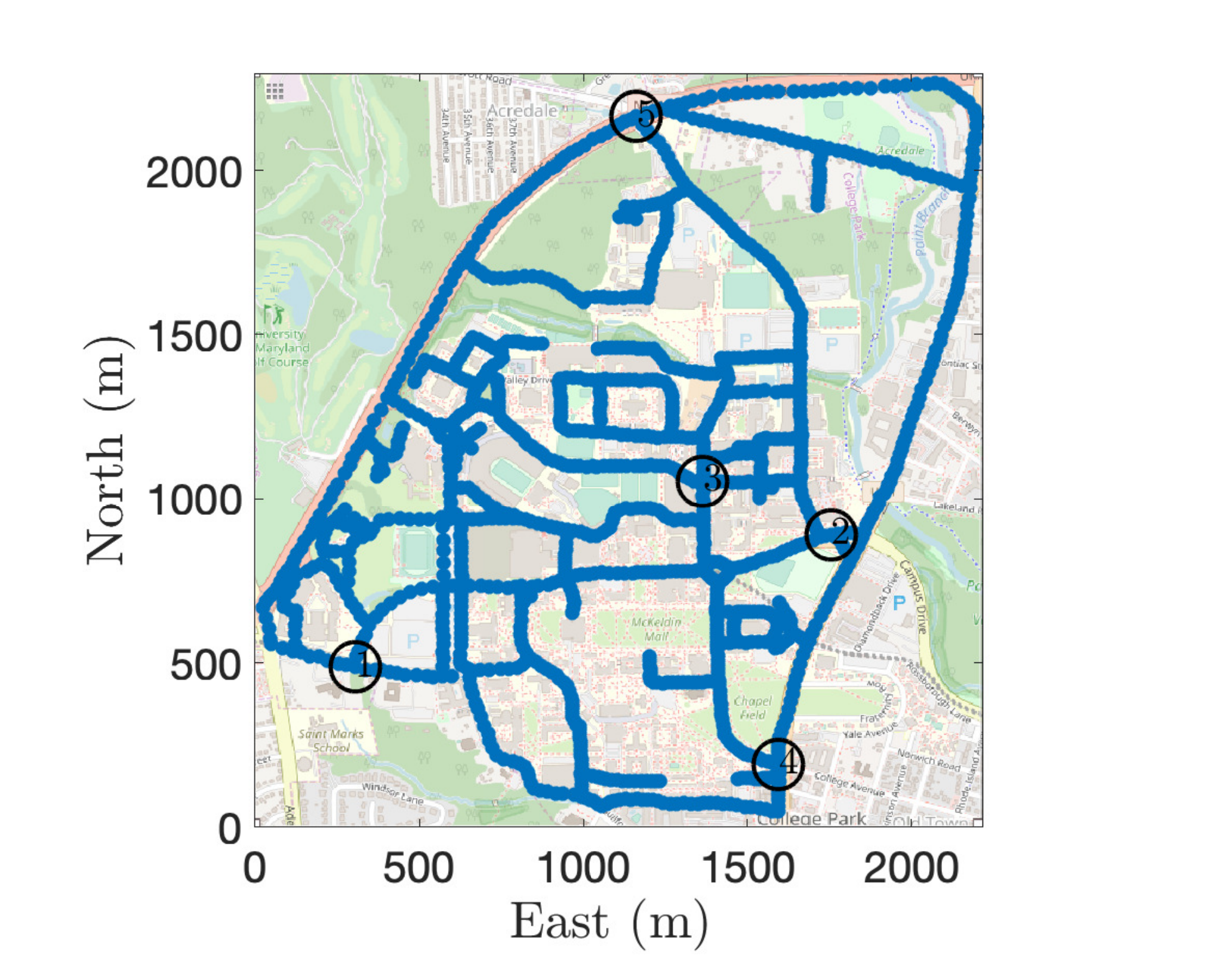}}
\subfloat[]{\includegraphics[trim=2cm 0cm 0cm 0cm, clip=true, width=0.25\textwidth]{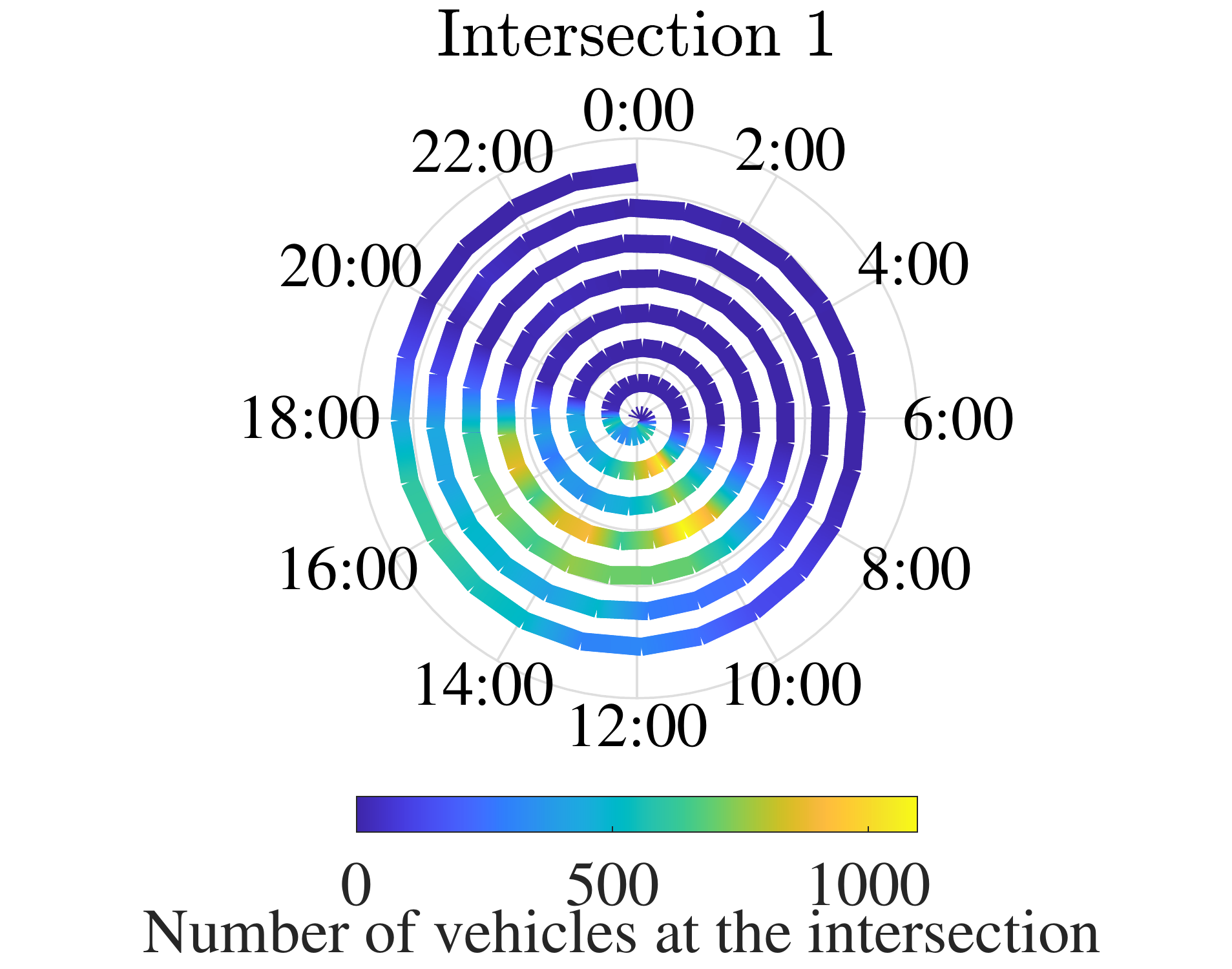}}
\caption{Schematic diagram of traffic data: (a) University of Maryland road network with Numina sensors, (b) Traffic congestion pattern of an intersection over a single week, each revolution denotes a day of the week with times marked as angles; the number of vehicles is denoted by the colormap. The daily pattern of peak congestion between mornings and afternoons is evident.} \label{Fig: TrafficSchematic}
\end{figure}

\begin{figure}[t]
\centering 
\subfloat[]{\includegraphics[trim=2cm 0cm 0cm 0cm, clip=true, width=0.5\textwidth]{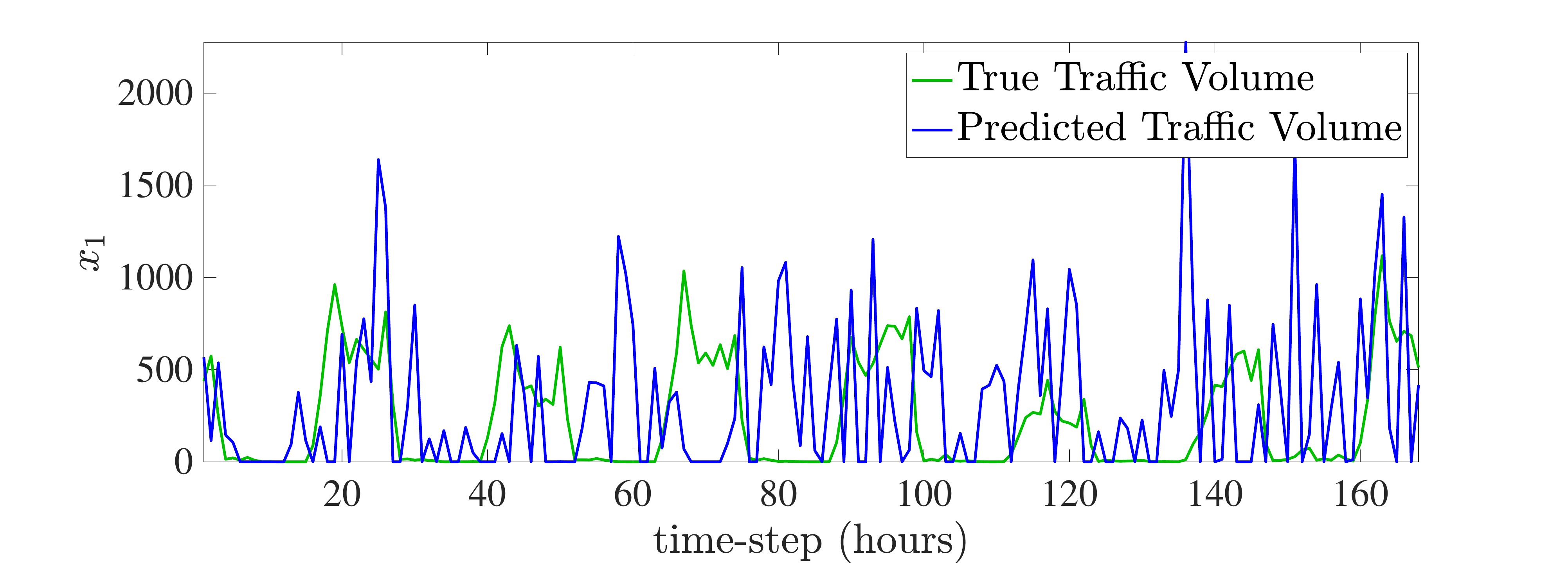}}\\
\subfloat[]{\includegraphics[trim=2cm 0cm 0cm 0cm, clip=true, width=0.5\textwidth]{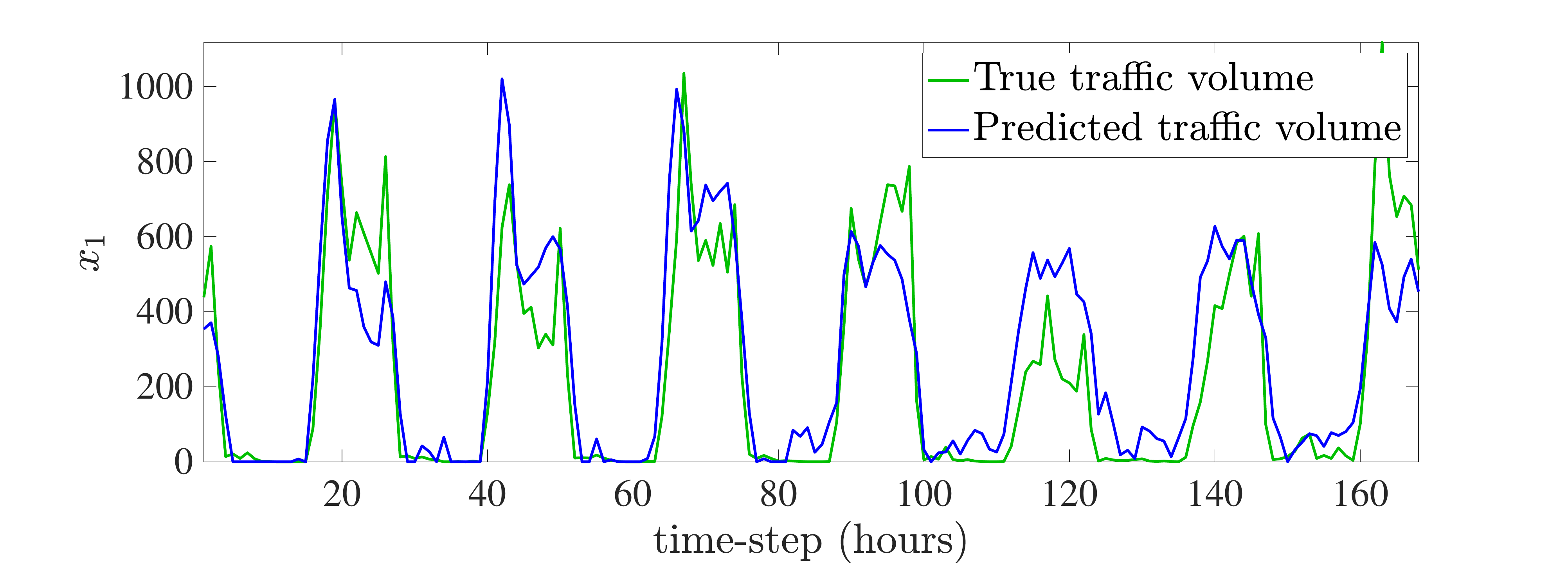}}
\caption{Estimation of the partially observed traffic-volume time-series recorded from the Numina sensor at intersection 1: (a) true and estimated traffic volume with 10 dimensional delay embedding ($m=10$), (b)  true and estimated traffic volume with 100 dimensional delay embedding ($m=100$)} \label{Fig: Numina}
\end{figure}

\begin{figure*}[t]
\centering 
\subfloat[]{\includegraphics[trim=0cm 0cm 0cm 0cm, clip=true, width=0.20\textwidth]{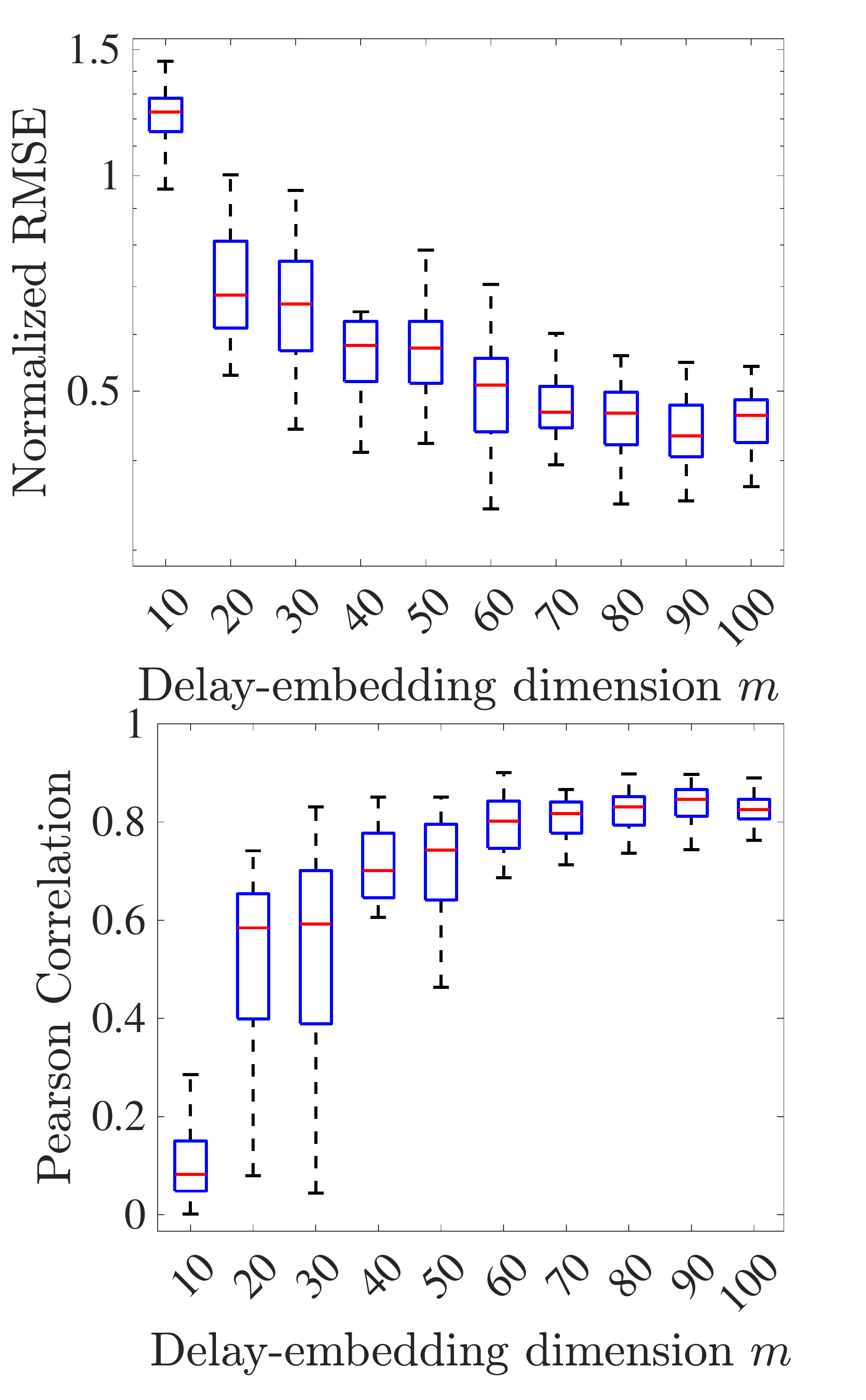}}
\subfloat[]{\includegraphics[trim=0cm 0cm 0cm 0cm, clip=true, width=0.20\textwidth]{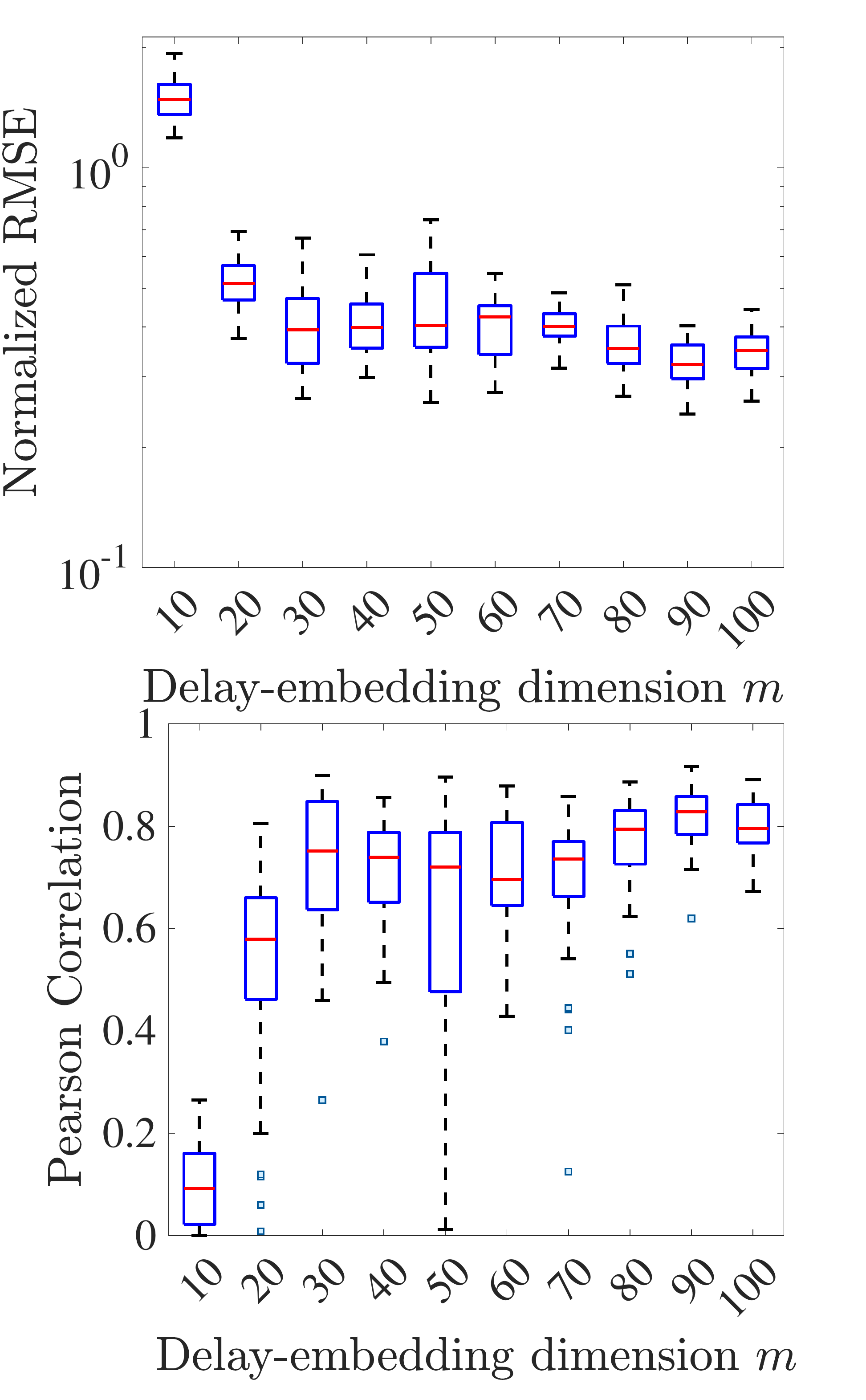}}
\subfloat[]{\includegraphics[trim=0cm 0cm 0cm 0cm, clip=true, width=0.20\textwidth]{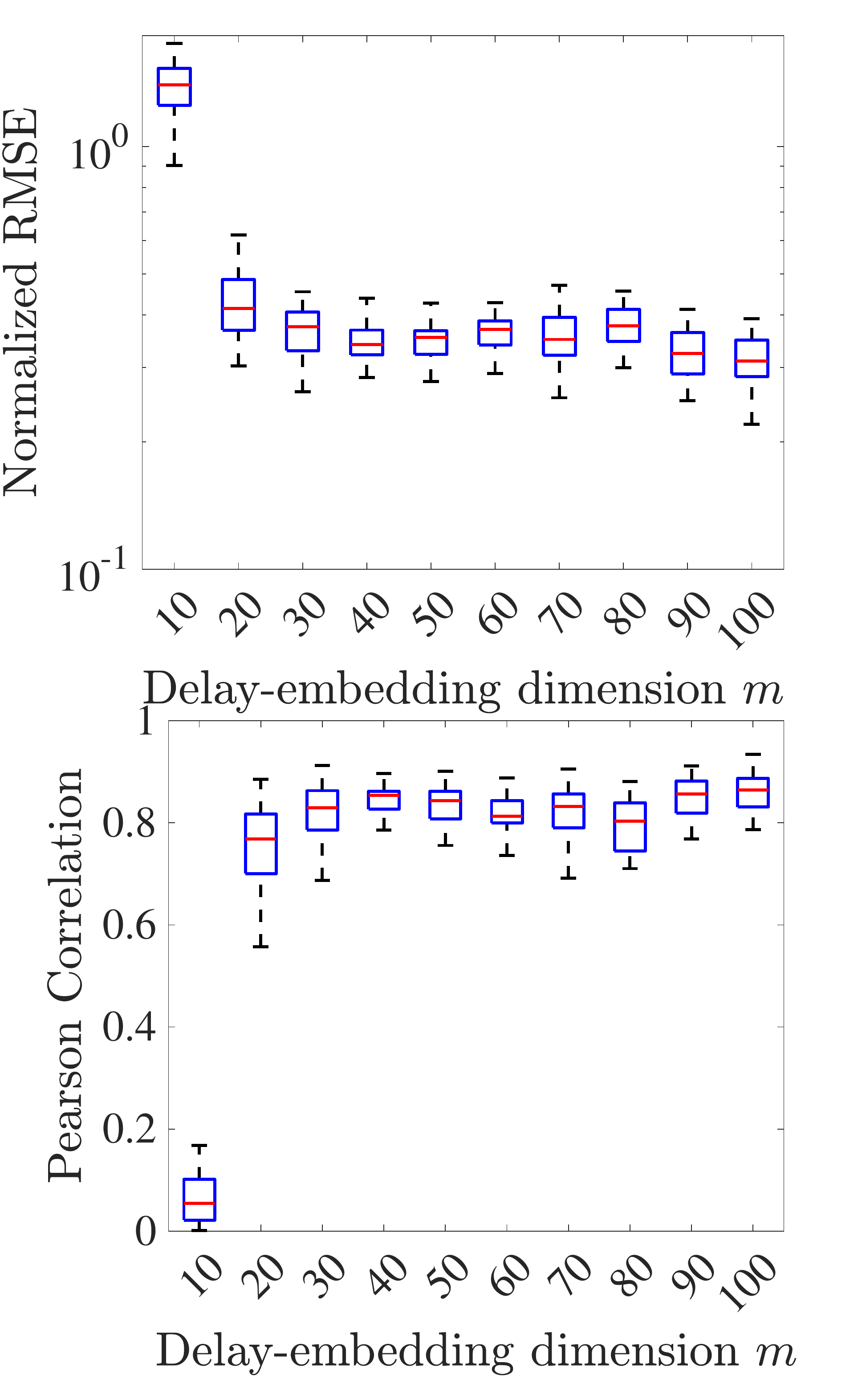}}
\subfloat[]{\includegraphics[trim=0cm 0cm 0cm 0cm, clip=true, width=0.20\textwidth]{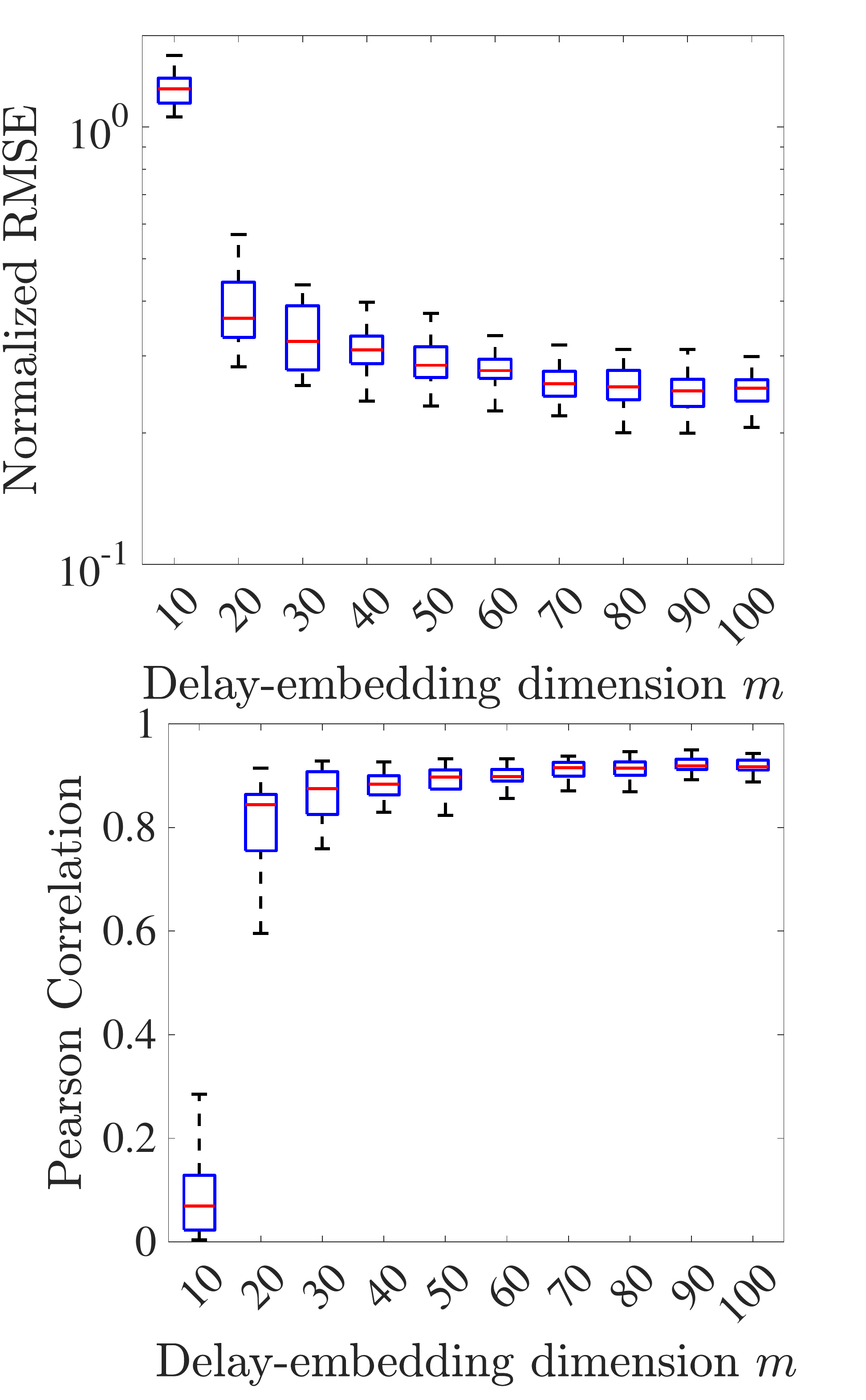}}
\subfloat[]{\includegraphics[trim=0cm 0cm 0cm 0cm, clip=true, width=0.20\textwidth]{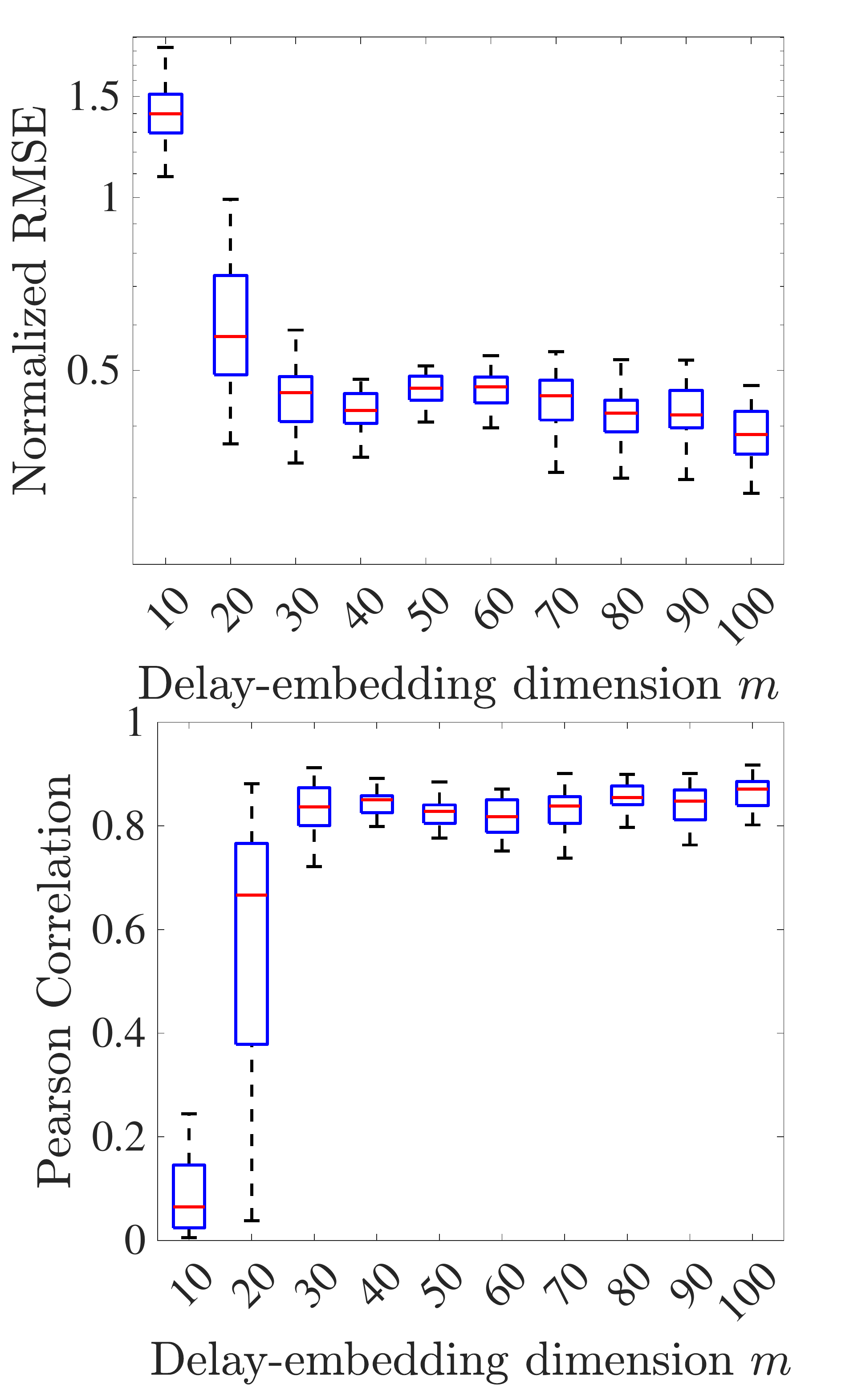}}
\caption{NRMSE and Pearson correlation coefficient with different embedding dimensions $m$: (a) Intersection 1, (b) Intersection 2, (c) Intersection 3, (d) Intersection 4, (e) Intersection 5} \label{Fig: NuminaError}
\end{figure*}

The proposed method is now applied to a dataset of traffic volumes obtained from Numina \citep{Numina} sensors at five different intersections on the University of Maryland campus. Fig.~\ref{Fig: TrafficSchematic}(a) represents the road network marked with sensor locations. Each sensor counts the number of pedestrians, bicycles, and vehicles at the respective intersections and store them in a server. We use the time series data of hourly vehicle traffic volume for two months. The ESN is trained on 1000 hours of traffic volume data and tested for one week, i.e., 168 hours. During each training, data from only one sensor is used. The training hyperparameters are listed in Table \ref{tb:hyparam}. Fig. \ref{Fig: Numina} shows the traffic volume prediction with delay embedding dimension of $m=10$ and $m=100$. Fig.~\ref{Fig: NuminaError} shows the NRMSE and Pearson correlation coefficient between predicted and true traffic volumes with sensor data from different intersections. The Pearson correlation coefficient between true and predicted sequences ($\{x(i): i=1,\ldots, l\}$ and $\{\hat{x}(i): i=1,\ldots, l\}$ respectively) measures their normalized linear correlation. It is given by
\bql
r(x,\hat{x}) = \frac{\sum\limits_{i}\left(x(i)-\bar{x}\right)^T\left(\hat{x}(i)-\bar{\hat{x}}\right)}{\sqrt{\sum\limits_{i}\norm{x(i)-\bar{x}}^{2}} \sqrt{\sum\limits_{i}\norm{\hat{x}(i)-\bar{\hat{x}}}^{2}}},
\eql
where $\bar{x}$ and $\bar{\hat{x}}$ denotes the time-average values of $x(i)$ and $\hat{x}(i)$.
 The performance of the ESN increases significantly with increasing embedding dimension $m$. 

\begin{remark}
This example proves that delay embedding can significantly improve an ESN's predictive power for quasiperiodic partial state data coming from a very high-dimensional system. Here, the traffic volume can be thought as a spatio-temporal dynamical system evolving over the road-network. 
\end{remark}

\section{Conclusion}
This paper describes a data-driven prediction method for partially observed systems and uses it to estimate the partial state measurements of three nonlinear systems from time-series data. The method utilizes the echo-state network (ESN) and Taken's embedding theorem for model identification using time-delay embedded partial state measurements. The prediction is carried out in a data-driven fashion without a dynamic model. The method is applied to a real data set of traffic patterns on the road network of the University of Maryland, College Park campus to predict the traffic volume at various intersections. In ongoing and future work, inference of unobserved states via time-delay embedded ESN with surrogate spatial interpolation model and a data-driven controller design will be investigated.

\begin{ack}
The author thanks Dr. Derek A. Paley and the University of Maryland Department of Transportation for the Numina sensor data. The author also thanks Dr. Artur Wolek for preprocessing the data.
\end{ack}

\bibliography{bibl}             
                                                   






\appendix
\section{What Happens with Unobservable Systems?} 
Consider the Lorenz system \eqref{Eq: Lorenz} with the observation $h([x, y, z]) = z$. This is an unobservable system since the system is invariant under the transformation $(x, y, z) \rightarrow (-x, -y, z)$, i.e., two different initial conditions on the strange attractor can produce the same observation data. In this case, it is shown in Fig.~\ref{Fig: LorenzError} that the delay embedding does not significantly improve the prediction accuracy of the ESN.

\begin{figure}[h]
\centering 
\subfloat[]{\includegraphics[trim=0cm 0cm 0cm 0cm, clip=true, width=0.25\textwidth]{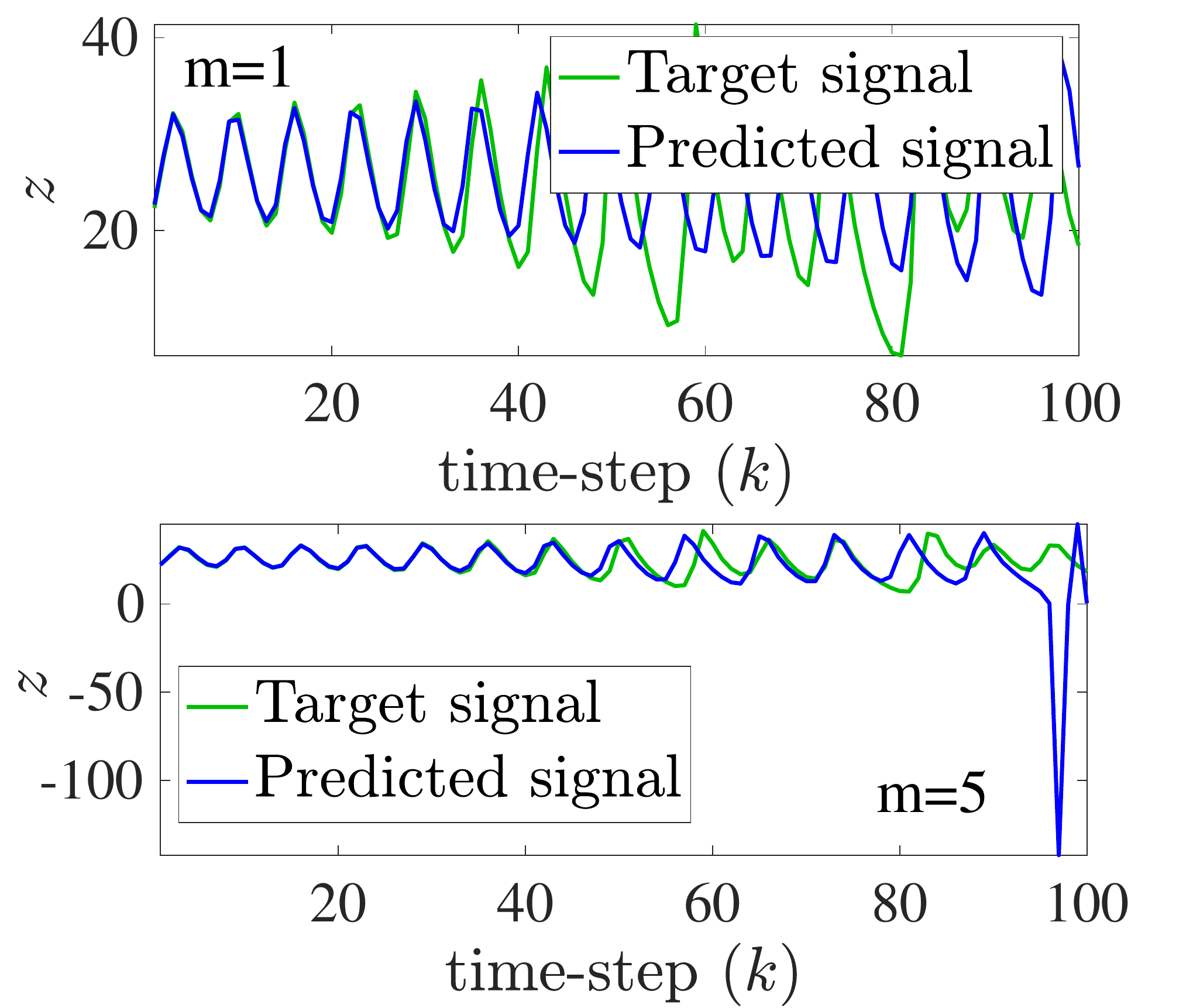}}
\subfloat[]{\includegraphics[trim=0cm 0cm 0cm 0cm, clip=true, width=0.25\textwidth]{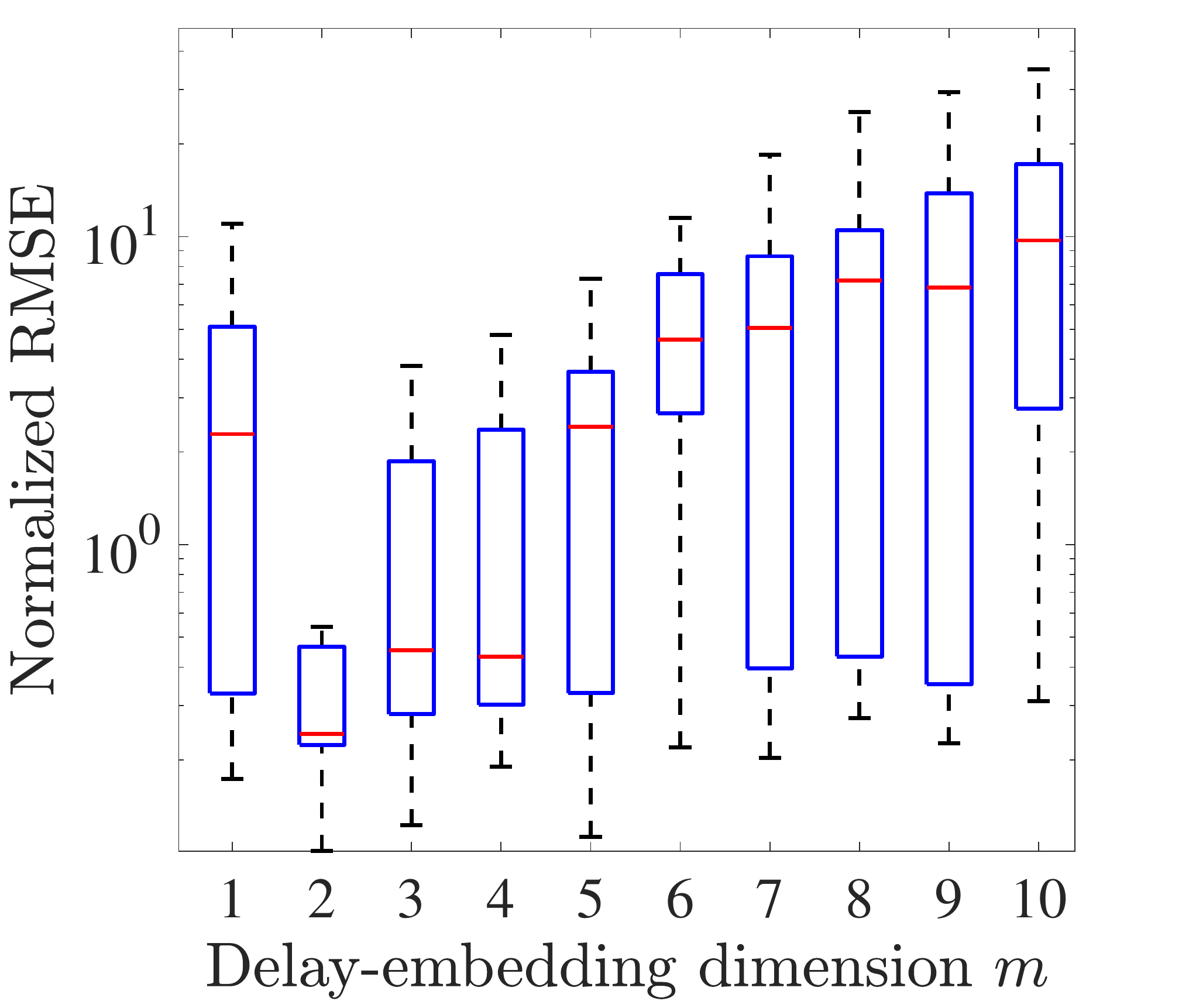}}
\caption{Estimation of the partially observed time-series $z(k)$ from Lorenz system \eqref{Eq: Lorenz}: (a) true and estimated signal with no delay embedding ($m=1$) and 5 dimensional delay embedding ($m=5$), (b)  NRMSE over 50 Monte-Carlo trials with different $m$. Delay embedding doesn't improve the performance over $m=2$ due to unobservability.} \label{Fig: LorenzErrorZ}
\end{figure}

\end{document}